\documentclass[prd,twocolumn,aps,noeprint,letterpaper]{revtex4-1}

\newcommand\x {\vec x}
\usepackage{amsmath,amssymb,bm,comment,color, graphicx, dsfont}
\usepackage{hyperref}
\allowdisplaybreaks

\newcommand{\ddd}{\displaystyle}

\newcommand{\dd}{{\rm d}}
\newcommand{\Tr}{{\mathrm{Tr}\,}}

\def\tR{{\tilde R}}
\def\tL{{\tilde L}}

\advance\voffset by 0.5cm

\makeatletter
\DeclareRobustCommand*{\bfseries}{%
  \not@math@alphabet\bfseries\mathbf
  \fontseries\bfdefault\selectfont
  \boldmath
}
\makeatother

\begin{document}

\title{On the temperature dependence of the chiral vortical effects}

\author{Tigran Kalaydzhyan}
\email{tigran.kalaydzhyan@stonybrook.edu}
\affiliation{Department of Physics and Astronomy, Stony Brook University,\\ Stony Brook, New York 11794-3800, USA}

\date{\today}
\begin{abstract}
We discuss the origins of temperature dependence of the axial vortical effect, i.e. generation of an axial current in a rotating chiral medium along the rotation axis. We show that the corresponding transport coefficient depends, in general, on  the number of light weakly interacting degrees of freedom, rather than on the gravitational anomaly. We also comment on the role of low-dimensional defects in the rotating medium, and appearance of the chiral vortical effect due to them.
\end{abstract}

\maketitle

\textbf{Introduction.} 
Quantum anomalies have recently attracted much attention due to their effect on the classical dynamics of chiral liquids.
Typical examples of such liquids are the strongly coupled quark gluon plasma, dense QCD at the CFL phase, superfluid $\mathrm{{}^3He-A}$, etc.
The anomaly effects can manifest themselves in the response of the fluid to an external magnetic field or rotation and can be studied through the transport coefficients in the vector (e.g., electric) or axial-vector currents,
\begin{align}
j^\alpha =& j^\alpha_{(0)} + \kappa_\omega \omega^\alpha + \kappa_{EM} B^\alpha + ...\,,\\
j^\alpha_5 =& j^\alpha_{5 (0)} + \xi_\omega \omega^\alpha + \xi_A B^\alpha + ...\,,\\
j^\alpha_{5B} =& j^\alpha_{5B (0)} + \xi_{\omega B} \omega^\alpha + \xi_{AB} B^\alpha + ...\,,\label{J5B}
\end{align}
where the $j_{(0)}$ denote the zero-order components (e.g., $j^\alpha_{(0)}=\rho u^\alpha$ for an ideal fluid, or $j^\alpha_{(0)}=\rho u^\alpha + \rho_S u^\alpha_S$ for a superfluid), $\omega^\alpha = \epsilon^{\alpha\beta\gamma\delta}u_\beta \partial_\gamma u_\delta$ and $B^\alpha = \epsilon^{\alpha\beta\gamma\delta}u_\beta F_{\gamma\delta}$ are the vorticity and the magnetic field, respectively, defined on the four-velocity of the liquid, $u_\alpha$. We also consider a \textit{baryon} axial current ($\ref{J5B}$) specific for the strong interactions. The ellipses denote higher-order corrections in derivative expansion.

Coefficient $\kappa_{EM}$ is the so-called chiral magnetic effect (CME) \cite{CME1}, $\kappa_\omega$ - chiral vortical effect (CVE) \cite{CVE}, $\xi_{A}$ - chiral separation effect (CSE) \cite{Vilenkin:1980fu,Metlitski,Son}, $\xi_\omega$ - axial vortical effect (AVE) \cite{Vilenkin:1979ui, Son:2004tq}. The latter one (and the similar coefficient in the baryon axial current) is the main focus of interest of this paper and usually has the form
\begin{align}
\xi_\omega = C(\mu^2 + \mu^2_5) + c_T T^2 + \mathcal{O}(\mu^2\mu_5,\mu_5^2,\mu\mu_5^2)\,,
\end{align}
where the first prefactor is the chiral anomaly coefficient, and $\mu$ and $\mu_5$ are the ordinary and the axial chemical potentials, respectively.
For the hydrodynamic derivation of these coefficients with one or several $\mathrm{U(1)_A}$, see, e.g., \cite{Son, Oz}, for $\mathrm{U(1)_V}\times\mathrm{U(1)_A}$ see \cite{Isachenkov,Kalaydzhyan:2011vx}, and for $\mathrm{U(N)_L}\times\mathrm{U(N)_R}$ \cite{Zahed, Lin:2011aa, Nair:2011mk}. In \cite{Oz} it was pointed out that there are temperature-dependent corrections to the anomalous coefficients, which appear as the integration constants and cannot be fixed from the hydrodynamics only. Similar coefficients appeared in \cite{CEE} in a context of the anomalous superfluidity and still could not be fixed. It was conjectured in \cite{Landsteiner:2011cp} that the coefficient $c_T$ originates from the mixed gauge-gravity anomaly coefficient, but, to our knowledge, there is no general proof of validity of this conjecture (the proportionality may, however, take place in some special circumstances \cite{Jensen:2012kj}, but the mixed anomaly itself depends on the microscopic properties of the
system \cite{Volovik:2003fe}). Moreover, our statement is that the coefficient $c_T$ is model dependent and reflects the statistical properties of chiral degrees of freedom in the system. This coefficient has been studied in the case of free rotating fermions \cite{Vilenkin:1980fu}, anomalous fermionic fluid \cite{Zahed:2012yu} and anomalous chiral superfluids \cite{Zahed:2013qia,Volovik:2003fe}
without introducing a mixed gauge-gravity anomaly. In all the cases the coefficient appeared as an integral over the Bose-Einstein or Fermi-Dirac distributions, which is a hint of microscopic, short-distance physics not captured (but allowed) by hydrodynamics.

To demonstrate a nonuniversal nature of the coefficient $c_T$ (and other temperature-dependent corrections) we choose a physical situation, where it changes within \emph{the same} model. We consider QCD with two massless flavors (at finite pion density) below and above the deconfinement transition. The system is subject to a uniform rotation. At low temperature the axial current will be carried by condensed $\pi^0$ mesons, while at high temperature - by chiral quarks.
We will show, that even the sign of temperature-dependent corrections is different in these two phases, due to the change of statistics for the chiral degrees of freedom.

As a final remark, even though the nonuniversal character of $c_T$ in relativistic hydrodynamics has attracted much attention recently, hints and examples of it are known for a long time in (nonrelativistic) condensed matter systems, see e.g. \cite{Volovik:2003fe, Salomaa:1987zz} and Refs. therein.

\textbf{Low temperatures.} 
In this section we consider QCD at low temperature, described by the chiral $N_f = 2$ Lagrangian with the gauged Wess-Zumino-Witten (WZW) term \cite{Wess:1971yu, Witten:1983tw} and nondynamical electromagnetic fields. We compute then the axial current and find $T^2$-corrections to AVE using the leading tadpole resummation technique \cite{ZahedBook}.
The action under consideration is given by \cite{Witten:1983tw}
\begin{align}
S\, =&\,  \frac{f_\pi^2}{4} \int\dd^4 x \,\,  \Tr[D_\alpha U^\dagger D^\alpha U]\label{WZW_action}\\
-&i \frac{N_c}{240 \pi^2}\int\dd^5 x \,\, \epsilon^{\alpha\beta\gamma\delta\zeta}\Tr[R_\alpha R_\beta R_\gamma R_\delta R_\zeta]\nonumber\\
-& \frac{N_c}{48 \pi^2}\int\dd^4 x \,\, \epsilon^{\alpha\beta\gamma\delta} A_\alpha \Tr[Q(L_\beta L_\gamma L_\delta  + R_\beta R_\gamma R_\delta)]\nonumber\\
+& \frac{i N_c}{24 \pi^2} \int\dd^4 x \,\, \tilde F^{\alpha\beta} A_\alpha \Tr[Q^2(L_\beta + R_\beta) + \nonumber\\
 &~~~~~~~~~~~ + \frac{1}{2}(Q U Q U^\dagger L_\beta + Q U^\dagger Q U R_\beta)]\,,\nonumber
\end{align}
with standard definitions
\begin{align}
R_\alpha \equiv U^\dagger \partial_\alpha U,\qquad & L_\alpha \equiv \partial_\alpha U U^\dagger,\label{LR_currents}\\
\tilde F^{\alpha\beta} \equiv \frac{1}{2} \epsilon^{\alpha\beta\gamma\delta}F_{\gamma\delta},\quad & D_\alpha \equiv \partial_\alpha + i e A_\alpha[Q, \cdot]\,.
\end{align}
Here the chiral fields $U$ can be represented as $U = \exp(i \pi^a \tau^a / f_\pi)$, with $\Tr(\tau^a\tau^b) = 2 \delta^{ab}$; $a, b=1,2,3$, and the charge matrix $Q = \mathrm{diag}(2/3, -1/3)=\mathds{1}/6 + \tau^3/2$. By means of the vector and axial transformations,
\begin{align}
U \xrightarrow[V]{}\, e^{i \varepsilon_V Q}\, U\, e^{-i \varepsilon_V Q}, \quad U \xrightarrow[A]{}\, e^{-i \varepsilon_A Q_5}\, U\, e^{-i \varepsilon_A Q_5}\,,
\end{align}
one can find a gauge-invariant and conserved vector (electric) current,
\begin{align}
j^\alpha =& i\frac{f_\pi^2}{2}\,\Tr[Q(\tR^\alpha - \tL^\alpha)] \label{vector_current}\\
            & - \frac{N_c}{48 \pi^2}\,\epsilon^{\alpha\beta\gamma\delta}\Tr[Q\tR_\beta \tR_\gamma \tR_\delta + Q\tL_\beta \tL_\gamma \tL_\delta]\nonumber\\
            &+ \frac{i N_c}{12 \pi^2}\, \tilde F^{\alpha\beta} \Tr[Q^2(\tL_\beta + \tR_\beta) \nonumber\\
            & ~~~ +\frac{1}{2}(QUQU^\dagger \tL_\beta + QU^\dagger QU \tR_\beta)]\,,\nonumber
\end{align}
as well as a gauge-dependent axial current,
\begin{align}
j_5^\alpha =& -i\frac{f_\pi^2}{2}\,\Tr[Q_5(\tR^\alpha + \tL^\alpha)] \label{axial_current}\\
            & + \frac{N_c}{48 \pi^2}\,\epsilon^{\alpha\beta\gamma\delta}\Tr[Q_5\tR_\beta \tR_\gamma \tR_\delta - Q_5\tL_\beta \tL_\gamma \tL_\delta]\nonumber\\
            &+ \frac{i N_c}{12 \pi^2}\, \tilde F^{\alpha\beta} \Tr[Q Q_5(\tL_\beta - \tR_\beta)] \nonumber\\
            &+\frac{N_c }{4\pi^2}\tilde F^{\alpha\beta}A_\beta \Tr[Q^2 Q_5]\,.\nonumber
\end{align}
Here $\tR_\alpha$ and $\tL_\alpha$ are (\ref{LR_currents}) with partial derivatives replaced by the covariant ones. One can redefine the current (\ref{axial_current}) such that the last (gauge-variant) term is moved to the right hand side of the the chiral anomaly expression,
\begin{align}
\partial_\alpha j_5^\alpha = -\frac{N_c}{4\pi^2} F_{\alpha\beta}\tilde F^{\alpha\beta} \Tr[Q^2 Q_5]\,.
\end{align}
Since we are interested in the condensed neutral pions, we can simplify the above currents (but not the Lagrangian)
by substituting $U = \exp(i \pi^3 \tau^3 / f_\pi)$ and $D_\alpha = \partial_\alpha$ (due to the structure of $Q$ or, physically, because of the electric neutrality of $\pi^0$) and using the identities
\begin{align}
R_{[\alpha} R_{\beta]} = - \partial_{[\alpha} R_{\beta]}+\ldots,\,\, L_{[\alpha} L_{\beta]} = \partial_{[\alpha} L_{\beta]}+\ldots\,,\label{MC_naive}
\end{align}
where ellipses will be important in the last section of this paper. The charge matrix $Q_5$ can be chosen differently, depending on which current we want to study, e.g. $Q_5=\tau^3/2$ for the usual axial current or $Q_5=\mathds{1}/3$ for the \textit{baryon} axial current, $j_{5B}$. Simplified currents take a form
\begin{align}
j^\alpha =& -\frac{N_c}{12\pi^2 f_\pi} \tilde F^{\alpha\beta}\partial_\beta \pi^3,\qquad j_{5}^\alpha = f_\pi \partial^\alpha \pi^3,\\
j_{5B}^\alpha =& \,\frac{N_c}{36\pi^2 f_\pi^2}\epsilon^{\alpha\beta\gamma\delta}\partial_\beta\pi^3\partial_\gamma \partial_\delta \pi^3\,. \label{j5_inpions}
\end{align}
At the next step we introduce the condensate velocity $u_S^\alpha$, from the condition that the zero-order term in the axial current becomes $j_5^{\alpha} = f_\pi \partial^\alpha \pi^3 \equiv j^{0}_5 u_S^\alpha \equiv \rho_5 u_S^\alpha$. Velocity depends on the (axial) chemical potential $\mu_5 \equiv \ddd\frac{\delta \mathcal{L}}{\delta \rho_5} = \frac{\partial_0 \pi^3}{f_\pi} = \frac{\rho_5}{f_\pi^2}$ and is, therefore, equal to
\begin{align}
u_S^\alpha = \frac{\partial^\alpha \pi^3}{f_\pi \mu_5}\,.\label{BEC_velocity}
\end{align}
This identification is typical for BEC and superfluids \cite{Fetter2008}. We add a subscript ``S'' to distinguish the condensate velocity from the velocity of the normal component $u^\alpha$. We assume the normal component to be absent at (or close to) zero temperature.
Using these definitions, one can rewrite the currents in a purely hydrodynamic form,
\begin{align}
j^\alpha =& -\frac{N_c}{12\pi^2} \mu_5 \tilde F^{\alpha\beta}u^S_\beta,\\
j_5^\alpha =& \rho_5 u_S^\alpha, \qquad j_{5B}^\alpha = \frac{N_c}{36\pi^2}\, \mu_5^2\, \omega_S^\alpha\,. \label{j5}
\end{align}
The vector current contains the CME term, while the baryon axial current contains the axial vortical term. There are also some comments in order. First, the vector current does not contain a zero-order part, $\rho u^\alpha$, because $\pi^0$ is neutral. Because of the same reason, there are no chiral vortical and chiral separation effects in this derivation (they are proportional to the ordinary $\mu=0$). Second, one would naively expect (\ref{j5_inpions}) and, hence, the vorticity $\omega_\alpha$ to vanish identically. However, in the presence of a nonvanishing total angular momentum the only way for the condensate to develop a vorticity is to have a singular $\pi^3$ field (since the condensate is, in general, curl free). This singularity has a nontrivial topology in the plane perpendicular to the vortex line (similar to \cite{Callan, Kirilin:2012mw}),
\begin{align}
[\partial^\perp_\alpha,\partial^\perp_\beta]\pi^a=2\pi f_\pi \delta^{(2)}(\x_\perp)\,,
\label{vortex_topology}
\end{align}
which can be derived from the Stoke's theorem. Property (\ref{vortex_topology}) makes the vorticity to be quantized and concentrated on a set of lines in this phase. It is important, that there is no possibility to create vorticity in this system just by quantum fluctuations, because the vorticity itself is a result of a nontrivial topology (\ref{vortex_topology}) protected against quantum corrections. Moreover, we emphasize the fact that the terms
of the form (\ref{j5_inpions})
 should be restored in all studies of the rotating relativistic superfluids.

After obtaining the transport coefficients at zero temperature, we can find thermal corrections to them for the temperature $T \ll f_\pi$. In order to do that, on should calculate expectation values of the currents and renormalize the pion fields and pion decay constant $f_\pi$ by considering tadpole diagrams coming from the quartic (in $\pi^a$) terms of the pion Lagrangian \cite{Pisarski:1997bq, Zahed}. Given that the pion loops in tadpoles are excited thermally with the Bose-Einstein occupation numbers, the renormalization constant for the pion fields, $\delta_{Z_\pi}=1-\pi_r/\pi$, will be proportional to
\begin{align}
\langle\pi^2\rangle_{T} \equiv \frac{\langle\vec \pi^2\rangle_{T}}{N_f^2 - 1} = \int \frac{2\pi \delta(p^2)}{\ddd e^{\omega/T}-1}\, \dd^4 p = \frac{T^2}{12}\,.
\end{align}
Expectation values of the currents can be found by contracting pairs of the pion fields in the cubic in $\pi$ expansions of the currents. The result is given by
\begin{align}
\langle j^\alpha \rangle=& -\frac{N_c}{12\pi^2 f_\pi} \tilde F^{\alpha\beta}\partial_\beta \pi^3 \left(1-\frac{4}{3}\frac{\langle\pi^2\rangle_{T}}{f_\pi^2} \right)\,,\label{jcorr}\\
\langle j_5^\alpha \rangle=& f_\pi \partial^\alpha \pi^3 \left(1-\frac{4}{3}\frac{\langle\pi^2\rangle_{T}}{f_\pi^2} \right)\,,\\
\langle j_{5B}^\alpha \rangle=& \frac{e N_c}{36\pi^2 f_\pi^2}\epsilon^{\alpha\beta\gamma\delta}\partial_\beta\pi^3\partial_\gamma \partial_\delta \pi^3\,.\label{j5corr}
\end{align}
The renormalization of $\pi^a$ and $f_\pi$ comes from the tadpole corrections to the pion propagator and the $\pi^0 \gamma \gamma$ vertex, which is well known (see \cite{Pisarski:1997bq} and Refs. therein),
\begin{align}
\pi^a_r = \pi^a \left(1 - \frac{T^2}{36 f_\pi^2}\right),\quad f_\pi^r = f_\pi \left(1 - \frac{T^2}{12 f_\pi^2} \right)\,.
\end{align}
Replacing the bare zero-temperature quantities by their renormalized values in (\ref{jcorr},\ref{j5corr}), we obtain
\begin{align}
\langle j^\alpha \rangle=& -\frac{N_c}{12\pi^2 f_\pi^r} \tilde F^{\alpha\beta}\partial_\beta \pi^3_r\left(1-\frac{T^2}{6 f_\pi^2} \right),\label{jcorr_r}\\
\langle j_{5B}^\alpha \rangle=& \frac{N_c}{36\pi^2 {f_\pi^r}^2}\epsilon^{\alpha\beta\gamma\delta}\partial_\beta\pi^3\partial_\gamma \partial_\delta \pi^3_r \left(1-\frac{T^2}{9 f_\pi^2} \right)\,.\label{j5corr_r}
\end{align}
In addition, the zero-order term in the axial current has the same form as it was before renormalization, $\langle j_5^\alpha \rangle= f_\pi^r \partial^\alpha \pi^3_r$, which does not change our identification of the fluid velocity, i.e. $j_5^\alpha (T) = \rho_5 u_S^\alpha$.
One can also immediately notice a modification of CME (\ref{jcorr_r}), which a priori is not protected against the \textit{thermal} corrections.
Turning back to the hydrodynamic formulation, we write down the final result,
\begin{align}
j^\alpha (T) =& -\frac{N_c}{12\pi^2} \mu_5 \tilde F^{\alpha\beta}u^S_\beta \left(1-\frac{T^2}{6 f_\pi^2} \right),\\
\quad j_{5B}^\alpha (T) =& \frac{N_c}{36\pi^2} \left(\mu_5^2 - \frac{\mu_5^2}{9 f_\pi^2} T^2 \right)\, \omega_S^\alpha\,. \label{j5_T}
\end{align}
The nature of temperature corrections in this phase is not related to the gravitational anomaly and is, actually, the same as for the temperature corrections to the chiral condensate \cite{Gasser:1986vb}. Since the coefficient in front of $T^2$ depends on $\mu_5$, we conclude that $c_T=0$ (unless there are additional circumstances, when $\mu_5$ cancels $f_\pi$). We should also mention that the reason we choose $\pi^0$ is that the other carriers of the axial charge, such as the $\eta$, $\eta'$ fields \cite{Son:2004tq} or the axionlike excitations of the quark gluon plasma \cite{Kalaydzhyan:2012ut}, are not renormalized by the tadpole resummation because the effective Lagrangian does not contain terms quartic in fields and quadratic in derivatives of these fields. If these fields were condensed instead of $\pi^0$, then the corrections in (\ref{j5_T}) vanish, which is one more piece of evidence supporting the model dependence of the $T^2$ coefficient.

\textbf{High temperatures.} As the system is heated, the fraction of the condensed phase becomes smaller, vanishing above the critical temperature. In the absence of dissipation, the total angular momentum should be transferred completely to the normal phase. Since the rotation is uniform,
the integrated (quantized) vorticity $\omega_S^\alpha$, defined on the condensate velocity $u_S^\alpha$,
at low-$T$ is the same as the integrated high-$T$ vorticity $\omega^\alpha$. At asymptotically high temperatures, the chirality is carried by free chiral fermions. We assume that there is some mechanism (probably of the topological nature), that translates the axial charge of the system at low $T$ to the imbalance between numbers of quarks with different chiralities at high $T$, characterized by the chiral chemical potential $\mu_5$. As before, we consider nondynamical gauge fields (otherwise, see \cite{Golkar:2012kb} for loop corrections). The transport coefficients can be computed from various derivatives of the grand thermodynamic potential \cite{CME1},
\begin{align}
\Omega = \sum\limits_{s=\pm}\int\frac{\dd^3 p}{(2\pi)^3}\left[ \omega_{p,s} + T\sum_{\pm}\log (1 + e^{-\frac{\omega_{p,s}\pm \mu}{T}}) \right]\nonumber\,,
\end{align}
where $\omega_{p,s}^2 = (p+s\mu_5)^2 + m^2$ and, since we are mostly interested in the vortical effects, we consider a weak magnetic field, $\sqrt{eB}<\mu_5$. For one quark flavor and color, the vector (axial) density is the derivative of $\Omega$ with respect to the vector (axial) chemical potential in the limit $m \to 0$. Taking into account this fact and some results of the Wigner function analysis \cite{Gao:2012ix}, we obtain
\begin{align}
&\kappa_{EM} = \frac{N_c}{4}\frac{\partial^3 \Omega}{\partial \mu^2\, \partial \mu_5}\Tr[Q^2]=\frac{N_c}{2\pi^2} \mu_5 \Tr[Q^2],\\
&\kappa_\omega = \frac{N_c}{2}\frac{\partial^2 \Omega}{\partial \mu\, \partial \mu_5}\Tr[Q]=\frac{N_c}{\pi^2}\mu\mu_5\Tr[Q],\\
&\xi_A = \frac{N_c}{4}\frac{\partial^3 \Omega}{\partial \mu^3}\Tr[Q]=\frac{N_c}{2\pi^2} \mu \Tr[Q],\\
&\xi_{\omega} = \frac{N_c N_f}{2}\frac{\partial^2 \Omega}{\partial \mu^2} = \frac{N_c N_f}{6}T^2 +\frac{N_c N_f}{2\pi^2}(\mu^2+\mu_5^2),\label{highTAVE}\\
&\xi_{AB} = \frac{1}{3}\cdot\frac{N_c}{4}\frac{\partial^3 \Omega}{\partial \mu^3}\Tr[Q]=\frac{N_c}{6\pi^2} \mu \Tr[Q],\\
&\xi_{\omega B} = \frac{1}{3}\cdot\frac{N_c}{2}\frac{\partial^2 \Omega}{\partial \mu^2} = \frac{N_c}{18}T^2 +
\frac{N_c}{6\pi^2}(\mu^2+\mu_5^2)\,.\label{highTAVEB}
\end{align}
As one can see, the $T^2$ coefficient changed sign compared to (\ref{j5_T}), which is due to the Fermi-Dirac statistics for the fermions. The reason why the $\mu$($\mu_5$)-independent $T^2$ coefficient is at all present in (\ref{highTAVE},\ref{highTAVEB}), in comparison to (\ref{j5_T}), is due to the properties of moments of the Fermi distribution $n_F(\omega)$,
\begin{align}
\int\limits_0^\infty \dd p\, p^n\, \sum\limits_\pm n_F(p \pm \mu) = \#\cdot T^{n+1} + \mathcal{O}(\mu^2T^{n-1})\,,
\end{align}
throughout the calculation of the transport coefficients. Growth of $c_T$ from zero at low temperatures to a nonvanishing constant at high temperatures is in a qualitative agreement with the quenched lattice data \cite{Braguta:2014gea}.

\textbf{Defects.} We are now addressing a subtle consequence of the identity (\ref{vortex_topology}), which is a modification of the Maurer–-Cartan equations (\ref{MC_naive}),
\begin{align}
R_{[\alpha} R_{\beta]} = - \partial_{[\alpha} R_{\beta]} + \sum_{i} i\pi \delta(x^\alpha_i)\delta(x^\beta_i) \tau^3 ,\label{ident1}\\
\quad L_{[\alpha} L_{\beta]} = \partial_{[\alpha} L_{\beta]} + \sum_{i} i\pi  \delta(x^\alpha_i)\delta(x^\beta_i) \tau^3\label{ident2}\,,
\end{align}
where we sum over vortices in the $\pi^0$ condensate, with vortex lines intersecting the $(\alpha, \beta)$ plane. Substitution of the identities (\ref{ident1},\ref{ident2}) into the WZW action (\ref{WZW_action}) will result in the appearance of two types of terms: the chiral/axial vortical terms, and terms of the two-dimensional action induced on the vortex line, $\mathcal{C}_i$, which are new. We focus on the latter one,
\begin{align}
S_{2D} = \frac{N_c}{36 \pi f_\pi}\int_{\mathcal{C}_i}\dd^2 x \,\, \epsilon^{\alpha\beta} (A_\alpha \partial_\beta\pi^3 - 3 f_\pi A_\alpha A_\beta)
\end{align}
As one can see, this is a source term for an electric current (not current density) along the string,
\begin{align}
j^{\alpha} = \frac{N_c \epsilon^{\alpha\beta}}{36 \pi f_\pi} (\partial_\beta\pi^3 - 6f_\pi A_\beta)\,,\label{2Dj}
\end{align}
which in the case of a pion condensate uniformly rotating along the $z$-axis is simply a persistent current,
\begin{align}
j^z = -\frac{N_c}{36 \pi f_\pi} \partial_t\pi^3 = -\frac{N_c \mu_5}{36 \pi}\,,\label{supercurrent}
\end{align}
along the (superconducting) vortex line. One can check that the electric currents on the string and in the bulk of the fluid are separately conserved in absence of external electric fields. Switching on an electric field along the string will result in a bulk (radial) electric current perpendicular to the string and the anomaly inflow \cite{Callan}.

Current (\ref{2Dj}) receives temperature corrections in a full analogy to (\ref{j5_inpions}),
\begin{align}
\langle j^{\alpha}\rangle =\,\frac{N_c \epsilon^{\alpha\beta}}{6 \pi} \left(\frac{\partial_\beta\pi_r^3}{6 f_{\pi}} \left(1 -\frac{T^2}{18 f_\pi^2} \right)- A_\beta \left(1 -\frac{T^2}{9 f_\pi^2} \right)\right).\nonumber
\end{align}
Considering, again, a simplified situation (\ref{supercurrent}) and taking into account that each vortex carries a quantum $\Omega_{\mathrm{quant}}=-2\pi/\mu_5$ of vorticity, we get temperature corrections to the total persistent current,
\begin{align}
\langle J_{\mathrm{tot}}^z \rangle = \frac{\langle j^z \rangle \cdot \langle\Omega^z\rangle}{\langle \Omega_{\mathrm{quant}}^z \rangle} = \frac{N_c \mu_5^2}{72 \pi^2} \left(1 -\frac{T^2}{9 f_\pi^2} \right)\langle\Omega^z\rangle\,.\label{supercurrent_ren}
\end{align}
This current is the superfluid version of the CVE, which was not considered before in the literature.

\textbf{Conclusions.} In this paper we have demonstrated the model dependence for the temperature corrections to the chiral/axial vortical effects, coming from the quantum statistics of the chiral degrees of freedom. In addition, we emphasized the importance of low-dimensional defects (singular fields) for the condensed systems in rotation. These defects give rise to a rich phenomenology, which will be considered in the future publications.

\textbf{Acknowledgements.} I would like to thank  Dmitri Kharzeev for proposing to study the topic and, especially, Aleksas Mazeliauskas for going through some of the calculations. I am also thankful to Karl Landsteiner, Ismail Zahed, Frasher Loshaj, Valentin Zakharov, Grigory Volovik, and Kristan Jensen for useful discussions. This work was supported in part by the U.S. Department of Energy under Contract DE-FG-88ER40388.


\end{document}